\documentclass[12pt]{article}
\usepackage{epsf,epsfig}
\usepackage{amssymb}
\usepackage{amsmath}
\usepackage{graphicx}
\usepackage{latexsym}

\def\lromn#1{\uppercase\expandafter{\romannumeral#1}}
\def\lromnl#1{\lowercase\expandafter{\romannumeral#1}}
\def\ds{\displaystyle}
\def\av#1{\langle #1 \rangle}

\begin{document}

\begin{titlepage}

\begin{center}

\hfill ICRR-Report-521-2005-4 \\
\hfill KEK-TH-1035 \\
\hfill STUPP-05-182 \\
\hfill TUM-HEP-601/05 \\
\hfill \today

\vspace{1cm}

{\large 
Relic Abundance
of LKP Dark Matter in UED model \\
including Effects of Second KK Resonances
}
\vspace{1cm}

{\bf Mitsuru Kakizaki}$^{a,b,\,}$\footnote{kakizaki@th.physik.uni-bonn.de},
{\bf Shigeki Matsumoto}$^{c,\,}$\footnote{smatsu@post.kek.jp},
{\bf Yoshio Sato}$^{d,e,\,}$\footnote{yoshio@krishna.th.phy.saitama-u.ac.jp} \\
and 
{\bf Masato Senami}$^{a,\,}$\footnote{senami@icrr.u-tokyo.ac.jp}
\vskip 0.15in
{\it
$^a${ICRR, University of Tokyo, Kashiwa 277-8582, Japan }\\
$^b${Physikalisches Institut der Universit\"at Bonn,
Nussallee 12, 53115 Bonn, Germany}\\
$^c${Theory Group, KEK, Oho 1-1, Tsukuba, Ibaraki 305-0801, Japan}\\
$^d${Department of Physics, Saitama University, Saitama 338-8570, Japan}\\
$^e${Physik-Department, Technische Universit\"at M\"unchen, \\ James-Franck-Strasse 85748 Garching, Germany}\\
}
\vskip 0.5in

\abstract{
We reevaluate the thermal relic density of the Kaluza-Klein (KK) 
dark matter in universal extra dimension models. 
In particular, we consider the effect
of the resonance caused by second KK particles on the density. 
We find that the
annihilation cross sections relevant to the density are significantly
enhanced due to the resonance when the Higgs boson mass is large enough
($m_h \gtrsim 200$ GeV). 
As a result, the mass of the dark matter particle consistent with
the WMAP observation is increased compared to 
the result which does not include any resonance.
}

\end{center}
\end{titlepage}
\setcounter{footnote}{0}

%
%

\vspace{1.0cm}
\lromn 1. \hspace{0.2cm} {\bf Introduction}
\vspace{0.5cm}

Recent cosmological observations have determined 
the cosmological parameters precisely. 
In particular, the large difference between the mean density of
the matter and the baryon has revealed that the large amount of
non-baryonic cold dark matter exists in our universe \cite{WMAP}. 
The present interest
concerning dark matter physics is its identification.

There are many discussions on the constituent of dark matter. 
Among those,
weakly interacting massive particles (WIMPs) 
are excellent candidates for dark matter. 
They can explain not only the relic
abundance of dark matter, 
but also the large scale structure of the present universe.

Many candidates for WIMPs have been proposed 
from models of particle physics.
One of the most attractive WIMP candidates is 
the lightest supersymmetric particle (LSP) in
supersymmetric (SUSY) extensions of the Standard Model (SM), 
and it has been extensively studied so far \cite{reviews}. 
Recently an alternative candidate for WIMPs \cite{LKP} 
has been proposed
in universal extra dimension (UED) models \cite{Appelquist:2000nn}, 
which are ones of well-motivated
scenarios with TeV-scale extra dimensions \cite{TeVXD}.

UED models are natural extensions of SM to higher space-time dimensions.
It is postulated that all particles in SM
propagate in the compact spatial extra dimensions. 
From the four-dimensional point of view, 
the models are described as SM with extra particle contents, 
which are the towers of the Kaluza-Klein (KK) particles 
associated with each SM particle. 
The KK mass spectra are quantized due to the compactification, 
and labeled by KK number $n$. 
In the case of five-dimensional space-time
the $n$-th KK particles have masses $m^{(n)}\sim n/R$, 
where $R$ is the size of the extra dimension. 
Momentum conservation along with the compact extra
dimension leads to the conservation of KK number.

The compactification is performed by an orbifold, 
which is required 
for reproducing the correct particle contents of SM.
In other words, 
it is needed for obtaining chiral fermions at the zeroth KK level. 
The orbifolding violates the conservation of KK number and 
leaves its remnant called KK-parity conservation. 
Under the parity, particles at even (odd)
KK levels have plus (minus) charge. 
As a result, the lightest first KK particle (LKP) is stabilized 
and is a viable candidate for WIMP dark matter. 
This situation is quite similar to conventional SUSY models,
in which the LSP is stabilized by R-parity.

The LKP dark matter is frequently discussed
\cite{Servant:2002aq}-\cite{Kakizaki:2005en}, 
especially in the light of the HEAT experiment,
which reported larger positron fraction in the cosmic rays 
than its expectation \cite{HEAT}.
The LKP has a possibility to account for the positron excess
from its annihilation in the galactic halo \cite{positron}.
On the other hand, 
it is difficult to explain the anomaly by 
annihilation of Majorana particles, 
such as LSP, due to the helicity suppression.

In this paper, 
we reevaluate the thermal abundance of the LKP dark matter in
the minimal UED model: 
the simplest UED scenario 
which extra dimension is compactified on an $S_1/Z_2$ orbifold.
The first KK mode of photon, $\gamma^{(1)}$, is the LKP in the setup. 
The full calculation of the annihilation cross section
at tree level relevant to the abundance has already been performed
\cite{Servant:2002aq}. 
On the other hand,
it was pointed out that one-loop diagrams 
in which second KK particles propagate 
in the $s$-channel significantly contribute to the cross sections
\cite{Kakizaki:2005en}.
Because the LKP dark matter
is non-relativistic at the freeze-out temperature, 
the incident energy of two
LKPs is almost equal to the masses of the second KK particles.
The one-loop diagrams show the resonant behavior.

We perform the full calculation of the relic abundance 
including the effects of these resonances. 
In the preceding study, we briefly discussed the effect
without considering the coannihilation processes \cite{Kakizaki:2005en}. 
On the other hand, as shown in Ref. \cite{Servant:2002aq},
the coannihilation effects between the LKP and first 
KK particles of right-handed charged leptons
significantly change the relic abundance of dark matter.
Therefore it is worth evaluating
how the relic abundance is affected by both 
the resonance and the coannihilation.
In fact we will show the amazing result that
the wide parameter region in the compactification scale,
$550$ GeV $\lesssim 1/R \lesssim$ $770$ GeV, 
is consistent with the WMAP observation when the
Higgs boson mass is large enough ($m_h \gtrsim 200$ GeV).

This paper is organized as follows. 
In the next section, 
we briefly review the minimal UED model.
In particular, we focus on the mass spectra of KK particles. 
The annihilation cross sections including the
effects of second KK resonances are discussed in Sec. \lromn 3. 
Using the cross sections, 
we calculate the relic abundance of the LKP dark matter and
evaluate the mass of dark matter consistent with the WMAP observation in
Sec. \lromn 4. 
Section \lromn 5 is devoted to summary and discussion.


\vspace{1.0cm}
\lromn 2. \hspace{0.2cm} {\bf Minimal UED model}
\vspace{0.5cm}

The simplest UED model called the ``minimal UED model'' postulates 
one additional extra dimension, 
which is compactified on an $S^1/Z_2$ orbifold. 
The compactification scale, 
the inverse of the size $R$ of the extra dimension is constrained by
the electroweak precision measurements as $1/R \gtrsim 300$ GeV 
\cite{Appelquist:2000nn,EWmeasurements}.

The particle contents in the minimal UED is the same as those of SM.
There are three gauge fields $G,W,B$ and one Higgs doublet $H$. 
The matter contents are three
generations of fermions: the quark doublets $Q_i$, 
the up- and down-type quark singlets $U_i$ and $D_i$, 
the lepton doublets $L_i$ and the charged lepton
singlets $E_i$. 
The Latin index $i$ runs over three generations.
All these fields are defined on the five-dimensional space-time.

From the four-dimensional point of view, 
we have the usual SM particles and their KK modes with identical charges.
Notice that all KK modes of the fermions are Dirac-type due to the vector
property of a five-dimensional model. 
The interactions between these
particles are completely determined by those of SM, 
and thus there is neither CP nor flavor problem in this model.

The UED model has some attractive features. 
One is the existence of a candidate particle for 
non-baryonic dark matter. 
Another is that the model is restrictive and
has few number of undetermined parameters. 
There appear only two parameters related to new physics, 
the compactification scale $1/R$ and the cutoff scale $\Lambda$,
which is usually taken to be $\Lambda R \sim {\cal O}(10)$
\cite{Appelquist:2000nn}.

Although the masses of KK particles at each KK level
are almost degenerate at tree level,
radiative corrections relax the degeneracy \cite{Cheng:2002iz}. 
Below we summarize radiatively corrected mass spectra related to our 
calculations.

As mentioned in the previous section,
the LKP is identified with the first KK mode of photon in the
minimal setup. 
The mass eigenstates and eigenvalues of the $n$-th KK photon,
$\gamma^{(n)}$, and $Z$ boson, $Z^{(n)}$, are obtained by
diagonalizing the mass squared matrix in the $(B^{(n)}, W^{3(n)})$ basis:
\begin{eqnarray}
  \left(
    \begin{array}{cc}
      \ds\frac{n^2}{R^2}
      +
      \ds\delta m^2_{B^{(n)}}
      +
      \ds\frac{g^{\prime 2}}{4}v^2
      &
      \ds\frac{g' g}{4}v^2
      \\
      \ds\frac{g' g}{4}v^2
      &
      \ds\frac{n^2}{R^2}
      +
      \ds\delta m^2_{W^{(n)}}
      +
      \ds\frac{g^2}{4}v^2
    \end{array}
  \right)~,
  \label{eq:LKP_mass_matrix}
\end{eqnarray}
where $g$ $(g')$ is the SU(2)$_L$ (U(1)$_Y$) gauge coupling constant, 
and $v \simeq 246$ GeV is
the vacuum expectation value of the Higgs field. 
The radiative corrections to the massive KK gauge bosons
are given by
\begin{eqnarray}
  \delta m^2_{B^{(n)}}
  &=&
  -\frac{39}{2}\frac{g^{\prime 2}\zeta(3)}{16\pi^4 R^2}
  -\frac{1}{6}\frac{g^{\prime 2}n^2}{16\pi^2 R^2}
  \ln \left(\Lambda^2 R^2\right)~,
  \nonumber \\
  \nonumber \\
  \delta m^2_{W^{(n)}}
  &=&
  -\frac{5}{2}\frac{g^2\zeta(3)}{16\pi^4 R^2}
  +\frac{15}{2}\frac{g^2n^2}{16\pi^2 R^2}
  \ln \left(\Lambda^2 R^2\right)~.
\end{eqnarray}
Given $1/R \gg v$,
$\delta m^2_{W^{(n)}} - \delta m^2_{B^{(n)}}$ exceeds
the off-diagonal elements in Eq. (\ref{eq:LKP_mass_matrix}).
The weak mixing angles are extremely small for KK modes,
and the LKP is dominantly
composed of the first KK mode of the hypercharge gauge boson.

The LKP is highly degenerate with the first KK modes of 
right-handed charged
leptons in mass even after including radiative corrections. 
The masses of the $n$-th KK lepton singlets are given by
\begin{eqnarray}
  m_{E^{(n)}}
  =
  \frac{n}{R}
  +
  \frac{9}{4}\frac{g^{\prime 2} n}{16\pi^2 R}
  \ln \left(\Lambda^2 R^2\right)~,
\end{eqnarray}
where we ignore the mass terms induced 
by the electroweak symmetry breaking,
because these terms are small enough compared to $1/R$. 
From the above equations,
the degeneracy is found to be $1$ \% level.

Let us turn to a discussion on the mass of 
the second KK mode of the neutral and CP-even
Higgs boson $h^{(2)}$ in detail. 
This is because the mass difference between this KK Higgs
and two LKPs plays an important role 
in the calculation of the annihilation
cross sections (see the next section).

\begin{figure}[t]
  \begin{center}
    \scalebox{0.8}{\includegraphics*{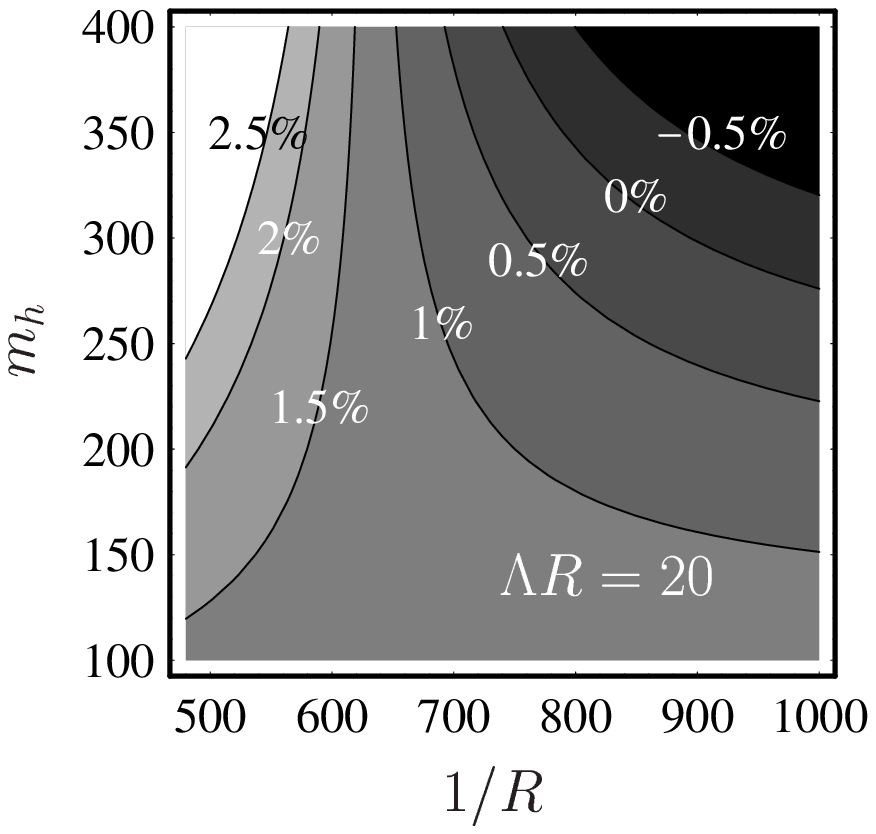}}
    \put(-95,-15){(a)}
    \hspace{0.5cm}
    \scalebox{0.8}{\includegraphics*{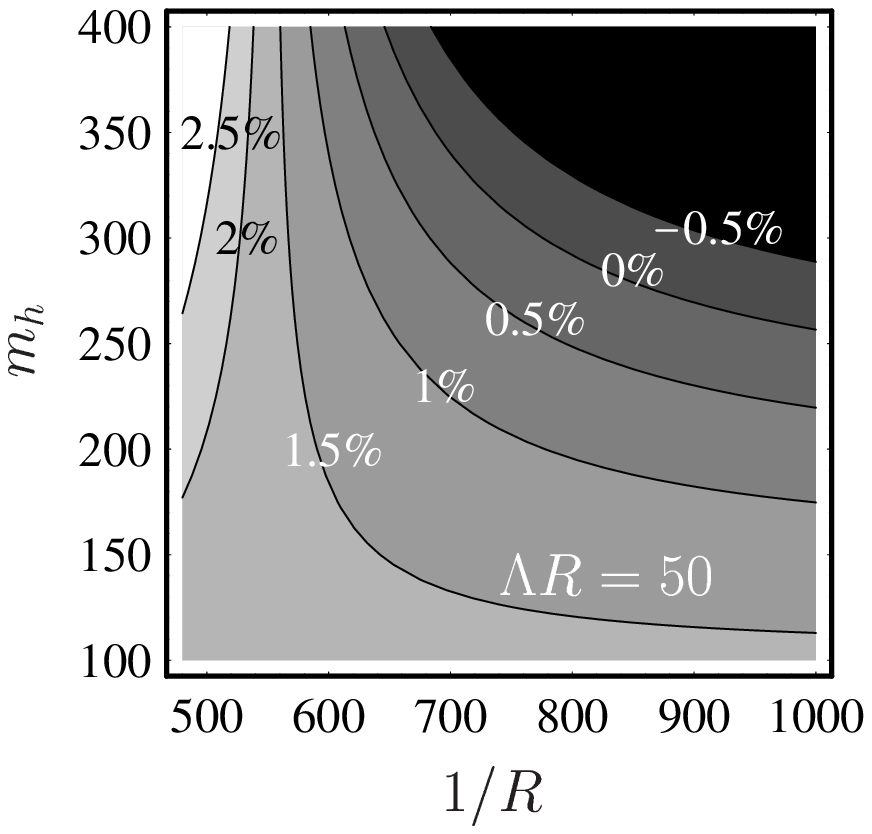}}
    \put(-95,-15){(b)}
    \caption{\footnotesize Contour plots of the mass splitting,
      $\delta \equiv (m_{h^{(2)}}-2m_{\gamma^{(1)}})/2m_{\gamma^{(1)}}$, 
      in the ($1/R$, $m_h$) plane for $\Lambda R = 20$ (a)
      and for $\Lambda R = 50$ (b).}
    \label{fig:degeneracy}
  \end{center}
\end{figure}

The mass of the neutral and CP-even Higgs boson is given by
\begin{eqnarray}
 m_{h^{(2)}}^2
 =
 m_h^2
 +
 \frac{4}{R^2}
 +
 \left(
  \frac{3}{2}g^2
  +
  \frac{3}{4}g^{\prime 2}
  -
  \frac{m_h^2}{v^2}
 \right)
 \frac{4}{16\pi^2 R^2}
 \ln \left(\Lambda^2 R^2\right)~,
\end{eqnarray}
where $m_h$ is the mass of the SM Higgs boson. 
The last term comes
from the radiative correction at one-loop level. 
In Fig. \ref{fig:degeneracy}, we show the contour plots of the mass 
splitting between $h^{(2)}$ and two $\gamma^{(1)}$s,
which is defined 
as $\delta \equiv (m_{h^{(2)}} - 2m_{\gamma^{(1)}})/2m_{\gamma^{(1)}}$,
in the ($1/R$, $m_h$) plane for
$\Lambda R = 20$ (a) and for $\Lambda R = 50$ (b). 
These figures show that the degeneracy is
${\cal O}$(1) \% level.
One also finds region where the mass difference is almost zero.


\vspace{1.0cm}
\lromn 3. \hspace{0.2cm}
{\bf Annihilation cross sections including second KK resonances}
\vspace{0.5cm}

In this section, 
we compute the annihilation cross sections including second
KK resonances in the self-annihilation of LKP dark matter (\lromnl 1), 
the coannihilation cross sections for the LKP and 
the degenerate particles in mass (\lromnl 2),
and the self-annihilation cross sections 
of the coannihilating particles (\lromnl 3). 
These cross sections
are required for calculating the relic abundance of LKP dark matter.

%
%

\vspace{1.0cm}
\underline{(\lromnl 1) Self-annihilation of LKP dark matter
($\gamma^{(1)}\gamma^{(1)}\rightarrow$ SM particles)}
\vspace{0.5cm}

The calculation of the annihilation cross section at tree level 
have already been performed in Ref. \cite{Servant:2002aq}:
\begin{eqnarray}
  \sigma^{({\rm Tree})}_{\rm Self}
  = \frac{95 \pi \alpha_{\rm em}^2}{81 \cos^4 \theta_W}
  \frac{5 (2m^2+s) L -7s\beta}{s^2 \beta^2} 
  + \frac{\pi \alpha_{\rm em}^2}{6 \cos^4 \theta_W s \beta}~,
   \quad
  L \equiv \ln\left(\frac{1 + \beta}{1 - \beta}\right) ~,
  \label{eq:BBL}
\end{eqnarray}
where $\alpha_{\rm em}$ is the fine structure constant,
$\theta_W$ is the weak mixing angle,
$m$ is the mass of the LKP, 
$s$ is the center of mass energy squared, 
$\beta$ is defined as $\beta^2 \equiv 1 - 4m^2/s$.
In the calculation, we assume that all first KK
modes have a equal mass and that all SM particles are massless.

At one-loop level, 
there is a diagram in which a second KK particle
propagates in the $s$-channel. 
The diagram plays an important role in 
the annihilation process in the early universe.
Since the incident particle is
non-relativistic when its interactions are frozen out, 
the $S$-wave component in the initial state dominantly
contributes to the process:
the CP-even $^1S_0$ and $^5S_2$ states.
Notice that the $^3S_1$ state is
forbidden due to the property of the neutral vector boson. 
Furthermore, the incident particle is electrically and color neutral.
Therefore the contributing second
KK particle is only the CP-even component of the second KK Higgs boson.

\begin{figure}[t]
  \begin{center}
    \scalebox{.92}{\includegraphics*{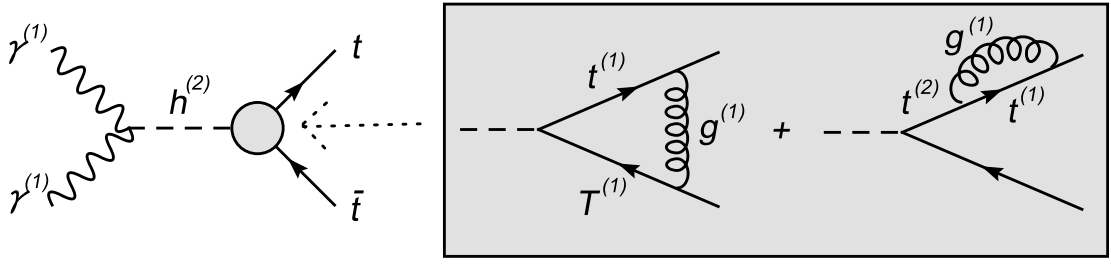}}
    \caption{\footnotesize Resonant annihilation process 
      of LKP dark matter through $s$-channel $h^{(2)}$.
      The dominant one-loop diagrams 
      to the $h^{(2)}-t-\bar{t}$ vertex stem
      from KK top quark--KK gluon mediation.
      Here $t$ is the zero mode of the top quark,
      and $t^{(n)}, T^{(n)}$ and $g^{(n)}$ are 
      the $n$-th KK modes of left- and right-handed top quarks and gluon
      respectively.}
    \label{fig:self_1_loop}
 \end{center}
\end{figure}

Although there is no transition from second KK states such as $h^{(2)}$
into states involving only SM particles at tree level,
their interactions appear through radiative corrections.
Among those, the dominant interactions between $h^{(2)}$
and SM particles come from the diagrams illustrated in Fig. 
\ref{fig:self_1_loop} (right two diagrams). 
After performing loop integrals and taking leading logarithmic parts, 
we have the following effective interaction:
\begin{eqnarray}
 {\cal L}_{\rm eff}
 =
 \frac{y_t\alpha_s}{12\pi}
 \ln\left(\Lambda^2R^2\right)
 h^{(2)}\bar{t}t~,
 \label{eq:h2-SM-SM}
\end{eqnarray}
where $y_t$ is the top Yukawa coupling constant and 
$\alpha_s$ is the strong gauge coupling constant.

Let us discuss the decay width of the second KK Higgs particle, 
that is important for calculating the annihilation cross section 
with the resonance.
The total decay width of $h^{(2)}$ is governed by the
decay mode into the top quarks through the one-loop diagrams
in Fig. \ref{fig:self_1_loop}.
In addition to the loop process,
there are several modes at tree level: 
the decay modes into two first KK particles and those
into one second KK particle and one SM particle. 
However, such tree-level processes are found to be forbidden or
highly suppressed due to a kinematical reason or small Yukawa couplings.
Therefore, 
the decay width of $h^{(2)}$ is dominated by the one-loop process,
and given by
\begin{eqnarray}
 \Gamma_{h^{(2)}}
 =
 \frac{y_t^2 \alpha_s^2 m_{h^{(2)}}}{384 \pi^3} 
 \left[
  \ln\left(\Lambda^2R^2 \right)
 \right]^2~.
 \label{eq:h2_width}
\end{eqnarray}

From the above effective interaction in Eq. (\ref{eq:h2-SM-SM}) 
and the width in Eq. (\ref{eq:h2_width}), 
we calculate the self-annihilation cross section
including the effect of the $h^{(2)}$ resonance \cite{Kakizaki:2005en}:
\begin{eqnarray}
  \sigma_{\rm Self} 
  &=&
  \sigma^{({\rm Tree})}_{\rm Self}
  +
 \sigma^{({\rm Res})}_{\rm Self}~,
 \nonumber \\
 \nonumber \\ 
 \sigma^{({\rm Res})}_{\rm Self}
 &=& \frac{\pi \alpha_{\rm em} \tan^2 \theta_W m_Z^2}{9 m \beta}
  \frac{\Gamma_{h^{(2)}}}
  {(s-m^2_{h^{(2)}})^2+m_{h^{(2)}}^2 \Gamma_{h^{(2)}}^2}
  \left(3 + \frac{s(s-4m^2)}{4m^4} \right)~.
\end{eqnarray}
Here,
$\sigma^{({\rm Tree})}_{\rm Self}$ is 
the tree-level result in Eq. (\ref{eq:BBL}),
while $\sigma^{({\rm Res})}_{\rm Self}$ is obtained by 
calculating the one-loop diagrams shown in Fig. \ref{fig:self_1_loop}. 
Notice that the interferential contribution between 
the tree-level diagrams and the one-loop diagrams is negligible, 
because it suffers from the chirality suppression of the top quark mass. 
Since the incident energy approximates the second KK Higgs mass,
$s \simeq (2m)^2 \simeq m^2_{h^{(2)}}$,
as emphasized in the previous sections, 
it is obvious that the resonant LKP dark matter annihilation 
is naturally realized.

\begin{figure}[t]
  \begin{center}
    \scalebox{0.9}{\includegraphics*{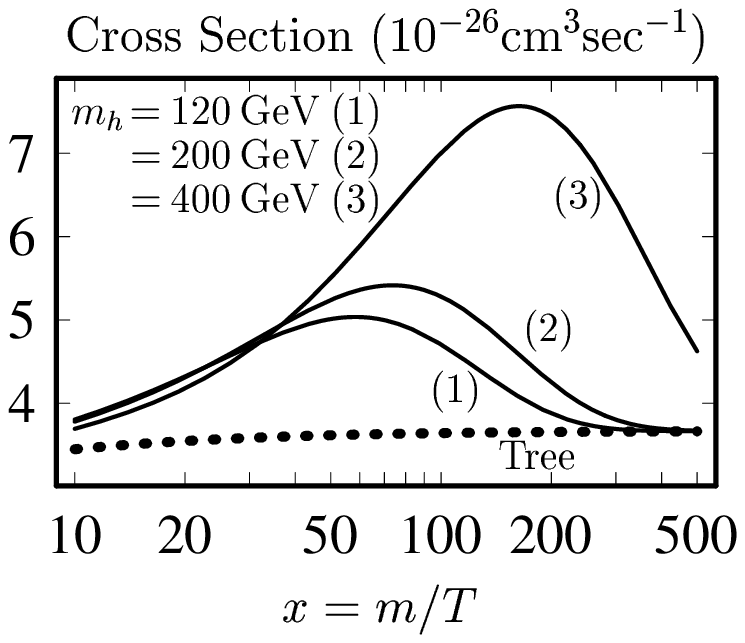}}
    \put(-100,-15){(a)}
    \hspace{0.7cm}
    \scalebox{0.9}{\includegraphics*{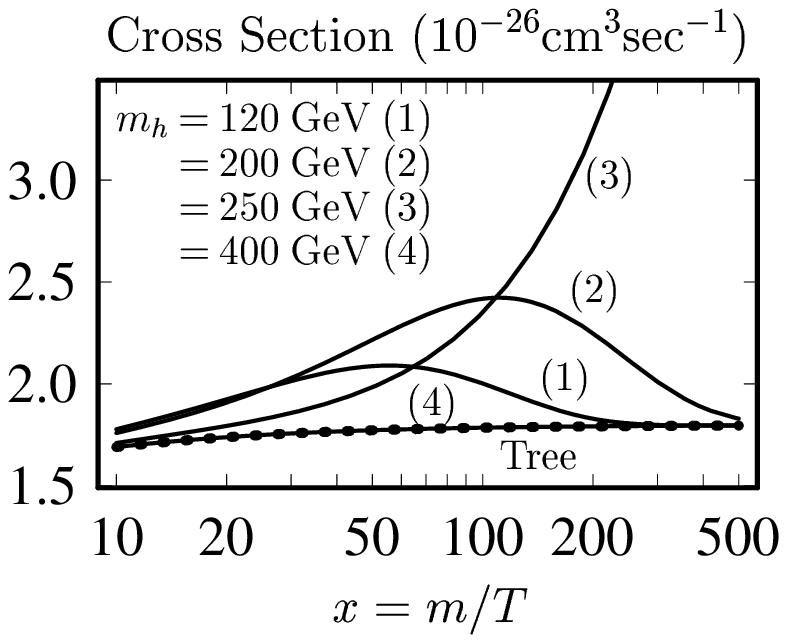}}
    \put(-100,-15){(b)}
    \caption{\footnotesize Thermally-averaged cross section
      as a function of the inverse of temperature $x = m/T$.
      Here we take $1/R = 700$ GeV and $\Lambda R = 20$ (a), and
      $1/R = 1000$ GeV and $\Lambda R = 50$ (b).
      In both Figs. (a) and (b), 
      the dependence of the thermally-averaged cross section 
      on the Higgs mass is shown.
      The dotted lines indicate the tree level results.
    }
    \label{fig:Self_CS}
  \end{center}
\end{figure}

The key quantity which controls the abundance
is the thermally-averaged annihilation cross section defined as
\begin{eqnarray}
 \av{\sigma v}_{\rm Self}
 =
 4\pi\left(\frac{m}{4\pi T}\right)^{3/2}
 \int_0^\infty dv \ 
 v^2
 (\sigma_{\rm Self} v)
 \exp\left(-\frac{mv^2}{4T}\right)~,
\end{eqnarray}
where $T$ is the temperature of the universe,
and the relative velocity is given by $v=2 \beta$.
At the rest of this paper, we use the notation $\av{\cdots}$ as
a thermal average.

We numerically performed the integration
and quantify the effect of the resonance on the annihilation process. 
In Fig. \ref{fig:Self_CS}, we show the averaged
cross section as a function of the inverse of the temperature $x = m/T$. 
The left figure (a) shows the result for the case of $1/R = 700$ GeV and
$\Lambda R = 20$, and three solid lines correspond to
$m_h=120$ GeV (1), $200$ GeV (2) and $400$ GeV (3).
For $m_h = 120$ GeV (1), the mass degeneracy between $h^{(2)}$ and two
$\gamma^{(1)}$s, 
$\delta = (m_{h^{(2)}} - 2m_{\gamma^{(1)}})/2m_{\gamma^{(1)}}$,
is $1.25$ \%, and $1.11$ \% for $m_h = 200$ GeV (2)
and $0.41$ \% for $m_h = 400$ GeV (3), respectively
(see Fig.\ref{fig:degeneracy}).
For comparison, the tree level calculation is also shown as a dotted line.

On the other hand, the right figure (b) is the result for $1/R = 1000$ GeV
and $\Lambda R = 50$, and four solid lines are
$m_h = 120$ GeV (1), $200$ GeV (2), $250$ GeV (3) and $400$ GeV (4).
The mass differences $\delta$ are $1.45$ \% , $0.73$ \%, $0.10$ \% and $- 2.72$
\% in these (1) to (4) cases.

The behavior and the magnitude of the thermally-averaged cross section
considerably vary depending on the parameters.
The annihilation cross section is enhanced due to the resonance compared to
the tree-level result when the mass difference $\delta$ is positive.
Smaller $\delta$ leads to larger enhancement at larger $x$.
In particular, $\delta$ is extremely small for the case (3) in the right figure,
and the peak of the cross section is out of the range of the figure.
In fact, the peak appears at $x \sim 3000$ and the cross section is increased to
$10^{-25}$cm$^3$sec$^{-1}$.
On the contrary, for the result of negative $\delta$ as in the case (4) in the
right figure, the resonance enhancement does not occur and the tree-level
cross section can be used for the calculation of the LKP abundance.

The enhancement of the cross section significantly contribute to the calculation
of the abundance as we will explicitly see in the next section. 

%
%

\vspace{1.0cm}
\underline{(\lromnl 2) Coannihilation
($\gamma^{(1)}E^{(1)}_i\rightarrow$ SM particles)}
\vspace{0.5cm}

Here, we discuss the coannihilation effects.
The LKP, $\gamma^{(1)}$, 
is found to be highly degenerate with the first KK modes of the
right-handed charged leptons, $E^{(1)}_i$,
as stated in the previous section.
It is thus indispensable to take the
coannihilation processes between these particles into account 
in calculating the LKP dark matter abundance.

The coannihilation cross sections at tree level are also calculated in
Ref. \cite{Servant:2002aq}:
\begin{eqnarray}
 \sigma^{({\rm Tree})}_{\rm Co}
 =
 \frac{\pi \alpha_{\rm em} ^2}{\cos \theta^4_W}
\frac{6L - \beta}{6 \beta^2 s}~,
\end{eqnarray}
where we assume that the mass of $E^{(1)}_i$ is equal to that of
$\gamma^{(1)}$.

\begin{figure}[t]
  \begin{center}
    \scalebox{0.9}{\includegraphics*{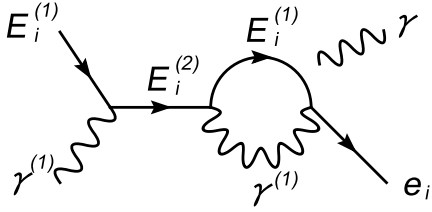}}
    \caption{\footnotesize Resonant one-loop diagram 
      which appears in the coannihilation process.}
    \label{fig:co_1-loop}
  \end{center}
\end{figure}

At one-loop level
there exist resonant processes in which the second KK modes
of the right-handed charged leptons are exchanged in the $s$-channel,
as shown in Fig. \ref{fig:co_1-loop}. 
The $E^{(2)}_i$-$e_i$-$\gamma$ vertex 
is given by dipole-type interaction,
which does not suffer from ultra-violet divergence
due to the gauge invariance.
These coannihilation processes, however, are found to be negligible.
The reason is as follows.
The second KK mode of the right-handed charged lepton, $E^{(2)}_i$, 
dominantly decays into $E^{(1)}_i$ and $\gamma^{(1)}$ at tree level. 
Although the process is kinematically suppressed, 
it is still very large compared to the
decay mode into $e_i$ and $\gamma$ at one-loop level. 
The branching ratio for
the one-loop process is indeed less than $0.1$ \%. 
As a result, the one-loop diagram
does not contribute to the coannihilation process. 
Therefore, we consider 
only the tree-level cross sections in the coannihilation processes.

%
%

\vspace{0.5cm}
\underline{(\lromnl 3) Self-annihilation of coannihilating particles
($E^{(1)}_iE^{(1)}_j\rightarrow$ SM particles)}
\vspace{1cm}

Finally, 
we consider the self-annihilation processes for three generations of 
the first KK lepton singlets $E^{(1)}_i$,
which are also important for calculating the LKP dark matter abundance. 
At tree level, the annihilation cross sections 
into SM particles are obtained as
\cite{Servant:2002aq}
\begin{eqnarray}
  && \sigma(E_i^{(1)} \bar{E}_i^{(1)}) 
  \nonumber \\
  && \quad =
  \frac{\pi \alpha^2_{\rm em}}{12 \cos^4 \theta_W m^2}
  \frac{s \beta (12 s^2 + 115 m^2 s + 68 m^4)
    - 12 L (2 m^2 s^2 - 5 m^4 s + 16 m^6)}{s^3 \beta^2}~, 
  \nonumber \\
  \nonumber \\
  && \sigma(E_i^{(1)} E_i^{(1)}) \nonumber \\
  && \quad = 
  \frac{\pi \alpha^2_{\rm em}}{2 \cos^4 \theta_W m^2}
  \frac{s \beta (2 s - m^2) + L m^2 (4 s - 5 m^2)}{s^2 \beta^2}~,
\end{eqnarray}
for the same lepton flavor, and, 
\begin{eqnarray}
 && \sigma (E_i^{(1)} \bar{E}_j^{(1)})
 = 
 \frac{\pi \alpha^2_{\rm em}}{4 \cos^4 \theta_W m^2}
  \frac{\beta(4 s + 9 m^2) - 8 L m^2}{s \beta^2}~,
  \nonumber \\
  \nonumber \\
 && \sigma (E_i^{(1)} E_j^{(1)})
 = 
 \frac{\pi \alpha^2_{\rm em}}{4 \cos^4 \theta_W m^2}
  \frac{4 s -3 m^2}{s \beta}~, 
  \quad i \ne j,
\end{eqnarray}
for different lepton flavor.
One might expect that there appear many resonances
via second KK particles in the $\bar{E}^{(1)}_i E^{(1)}_i$ 
annihilation processes.
However, this remark is not applicable to the minimal UED model.
The reason is as follows. 
In the processes, 
the initial two-body state is electrically and color singlet, 
and $E^{(1)}_i$ is singlet under the SU(2)$_L$ gauge group.
Furthermore only the $S$-wave state is relevant to
non-relativistic annihilation. 
As a result, possible candidates for second KK particles
causing resonant annihilation are the second KK mode of 
neutral pseudoscalar Higgs boson, $A^{(2)}$, 
and that of photon, $\gamma^{(2)}$. 
However both particles
do not contribute to the annihilation cross section. 
The diagram in which
$A^{(2)}$ propagates in the $s$-channel is strongly suppressed due to 
the small Yukawa coupling. 
As for the $\gamma^{(2)}$-mediated process, 
the mass of $\gamma^{(2)}$ is always smaller 
than twice the mass of $E^{(1)}_i$,
avoiding any resonance.
Since we can safely neglect any second KK resonance in these processes,
the tree-level cross sections are sufficient for our calculation.


\vspace{1.0cm}
\lromn 4 \hspace{0.2cm} {\bf LKP dark matter abundance}
\vspace{0.5cm}

We now evaluate the thermal relic abundance of the LKP dark matter 
using the annihilation cross sections obtained 
in the previous section and explore
the parameter region of the model consistent with the WMAP observation.
Since three generations of the next to LKPs, $E_i^{(1)}$, are
highly degenerate with the LKP in mass,
they are left over at the decoupling and decay after then.
In this sense,
$E_i^{(1)}$ should be also recognized as dark matter particles
in the early universe.
Hence,
the present density of dark matter is given by 
the sum of the number densities of the LKP and the next to LKPs.

The procedure for the calculation was developed 
in Ref. \cite{Griest:1990kh}.
The time evolution of the total number density of dark matter,
$n \equiv n( \gamma^{(1)}) + \sum_i n (E_i^{(1)})$,     
obeys the following Boltzmann equation:
\begin{eqnarray}
 \frac{dY}{dx}
 =
 -\frac{\langle \sigma v \rangle_{\rm eff}}{Hx}s
 \left(
  Y^2 - Y_{\rm eq}^2
 \right)~,
 \qquad
 Y_{\rm eq}
 = 0.145 \frac{g_{\rm eff}}{g_{*}} x^{3/2} e^{-x}
 ~,
 \label{eq:BE}
\end{eqnarray}where 
\begin{eqnarray}
  g_{\rm eff} 
  = 3 + 12 (1 + \Delta)^{3/2} e^{-x \Delta}, \quad
  \Delta = (m_{E^{(1)}} - m_{\gamma^{(1)}})/m_{\gamma^{(1)}}.
\end{eqnarray}
Here $Y = n/s$ ($Y_{\rm eq} = n_{\rm eq}/s$) describes the
number density $n$ ($n_{\rm eq}$) divided by the entropy density $s$
of the universe,
and $x = m/T$ parametrizes the inverse of the temperature.
The entropy density is given by
$s = 0.439\ g_* m^3/x^3$,
with $g_* $ being the number of relativistic degrees of freedom. 
The present entropy density is $s_0=2900$ cm$^{-3}$.
The Hubble parameter is
$H = 1.66\ g_*^{1/2}m^2/x^2 m_{\rm Pl}$, 
where $m_{\rm Pl} = 1.22\times 10^{19}$ GeV is the Planck mass scale. 

The effective cross section $\sigma_{\rm eff}$ involves not only the 
self-annihilation of LKP dark matters but also coannihilations,
and given by
\begin{eqnarray}
  g_{\rm eff}^2 \sigma_{\rm eff} 
  & = & 
  9 \ \sigma_{\rm Self}
  + 72 (1 + \Delta)^{3/2} e^{- x \Delta} \
  \sigma_{\rm Co}^{\rm (Tree)}
  \nonumber \\
  && + 24 ( 1 + \Delta)^3 e^{-2 x \Delta}\
  \left[ \sigma \left( E_i^{(1)}E_i^{(1)} \right)
    + \sigma \left( E_i^{(1)} \bar{E}_i^{(1)} \right) \right]
  \nonumber \\
  && + 48 ( 1 + \Delta)^3 e^{-2 x \Delta}\
  \left[ \sigma \left( E_i^{(1)}E_j^{(1)} \right)
    + \sigma \left( E_i^{(1)} \bar{E}_j^{(1)} \right) \right], 
  \quad i \ne j.
\end{eqnarray}
As the mass of a coannihilating particle is heavy, 
the effect of the coannihilation on the relic abundance
is exponentially suppressed.

By solving the Boltzmann equation in Eq. (\ref{eq:BE}),
we obtain the present abundance of dark matter $Y_\infty$. 
Since the non-relativistic LKP annihilation rate 
is enhanced by the resonance,
numerical calculation is required in order to obtain reliable results
\cite{Griest:1990kh}.
It is useful to express the relic density in terms of
$\Omega h^2 = m n h^2/\rho_c$, 
which is the ratio of the dark matter energy
density to the critical density in the present universe,
$\rho_c = 1.1 \times 10^{-5} h^2\ {\rm cm}^{-3}$.
The small letter $h$ denotes the scaled Hubble parameter and
takes a value $h = 0.71^{+0.04}_{-0.03}$.

\begin{figure}[t]
  \begin{center}
    \scalebox{0.9}{\includegraphics*{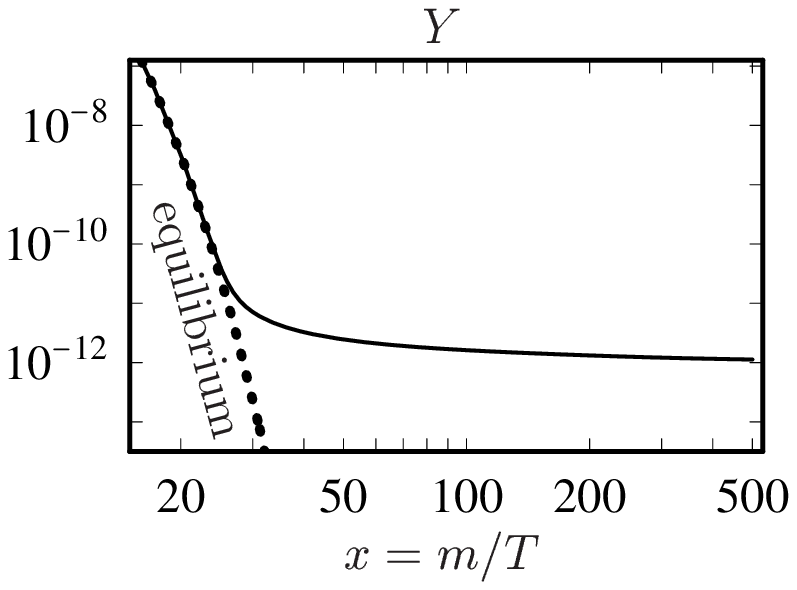}}
    \put(-100,-15){(a)}
    \hspace{0.7cm}
    \scalebox{0.86}{\includegraphics*{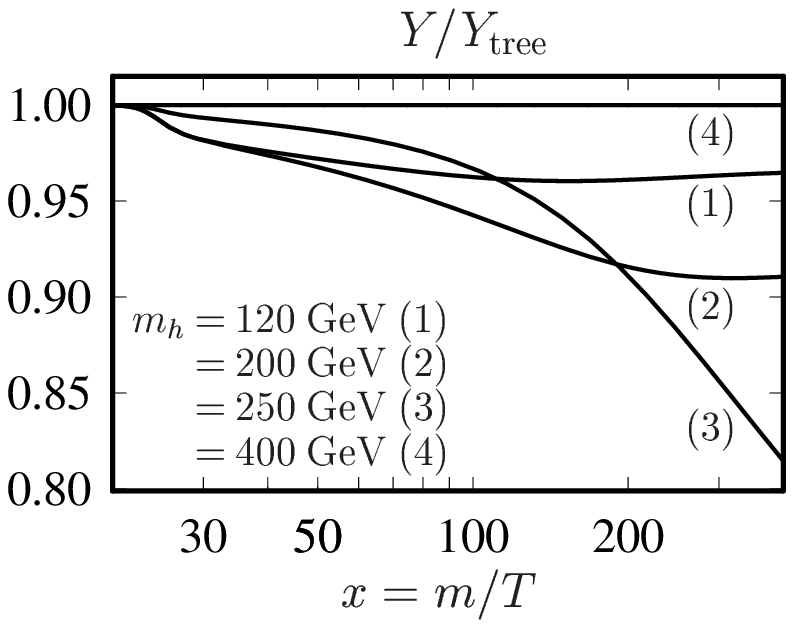}}
    \put(-100,-15){(b)}
    \caption{\footnotesize (a) Typical evolution of the abundance of 
      LKP dark matter as a function of the temperature of the universe 
      $x = m/T$. 
      The dotted line indicates the abundance of the LKP 
      at thermal equilibrium.
      (b) Ratio of the abundance $Y$ including 
      the effect of the second KK
      resonance to that at the tree level $Y_{\rm tree}$
      after the freeze-out $(20 < x < 400)$. 
      Here we take $1/R = 1000$ GeV and $\Lambda R = 50$.
      Four lines correspond to
      the cases, $m_h = 120$ GeV (1), $200$ GeV (2), 
      $250$ GeV (3)
      and $400$ GeV (4).}
    \label{fig:T-evolve}
  \end{center}
\end{figure}

It is instructive to observe a typical evolution
of the abundance, which is shown in Fig. \ref{fig:T-evolve}(a).
In the early universe, the abundance is
evolved on the line of the thermal equilibrium. 
After the interaction rate 
$\Gamma =  n \langle \sigma v \rangle_{\rm eff}$ 
drops below the expansion rate $H$ of the universe,
the dark matter particles depart from the equilibrium.
It is clear from Fig. \ref{fig:T-evolve}(a) that the reactions
are frozen out around $T \sim m/25$.

Generically,
a present abundance is almost determined 
around the freeze-out temperature.
However, 
the statement is not applicable to the case where 
the resonant annihilation occurs.
After the freeze-out, 
the averaged annihilation cross section is significantly enhanced 
depending on temperature, 
as discussed in the previous section.
The number density of dark matter particle gradually decreases
due to the enhancement of the cross section,
and eventually a smaller number of relics are left over,
even if the abundance is large at the freeze-out temperature.

The phenomenon of the ``late time decreasing'' is 
clearly seen in Fig. \ref{fig:T-evolve}(b).
Here we plot the ratio of the abundance $Y$ to that at the tree level
$Y_{\rm tree}$ as a function of $x = m/T$ 
after the freeze-out $(20 < x < 400)$.
We set the compactification scale to be $1/R = 1000$ GeV and
the cutoff scale to be $\Lambda R = 50$. 
Four lines correspond to the cases when $m_h = 120$ GeV (1), 
$200$ GeV (2), $250$ GeV (3) and $400$ GeV (4).
This figure demonstrates the decrease of the abundances 
due to the $h^{(2)}$ pole.
Especially for $250$ GeV (3),
the mass difference between two LKPs
and the second KK Higgs boson is so tiny
that the dark matter particles are found to be efficiently annihilated. 
(In the case of (3) the late time decreasing stops at $x \sim 3000$.)

\begin{figure}[t]
  \begin{center}
    \scalebox{0.80}{\includegraphics*{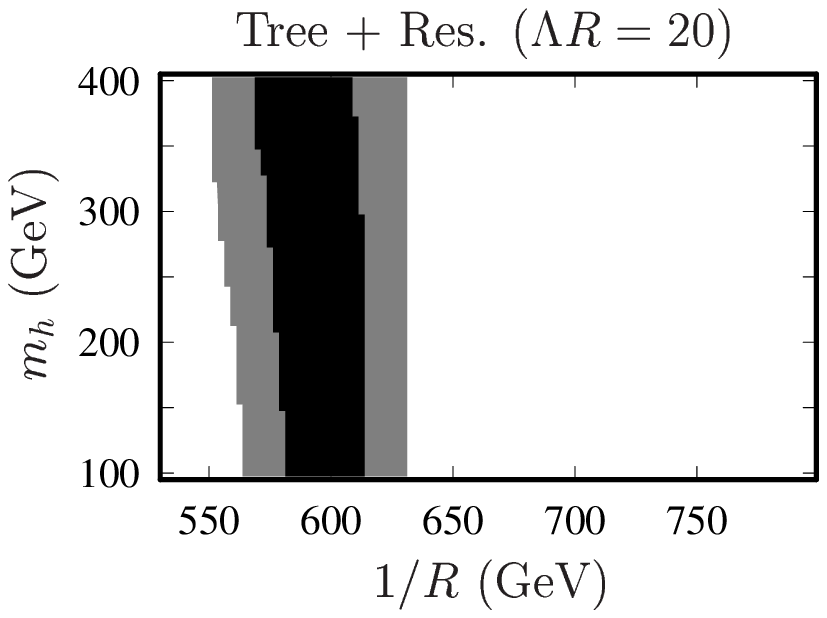}}
    \put(-100,-15){(a)}
    \hspace{0.5cm}
    \scalebox{0.80}{\includegraphics*{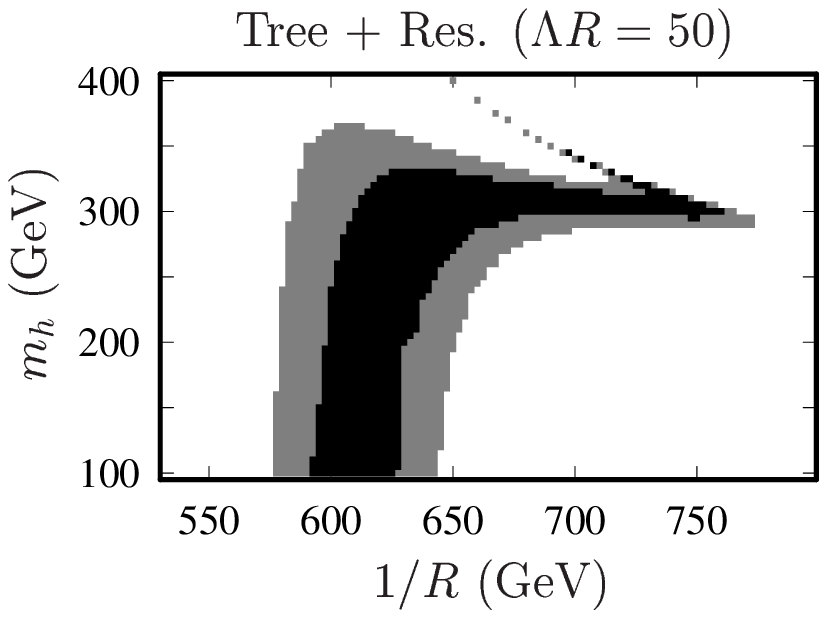}}
    \put(-100,-15){(b)}
    \vspace{0.5cm}
    \scalebox{0.80}{\includegraphics*{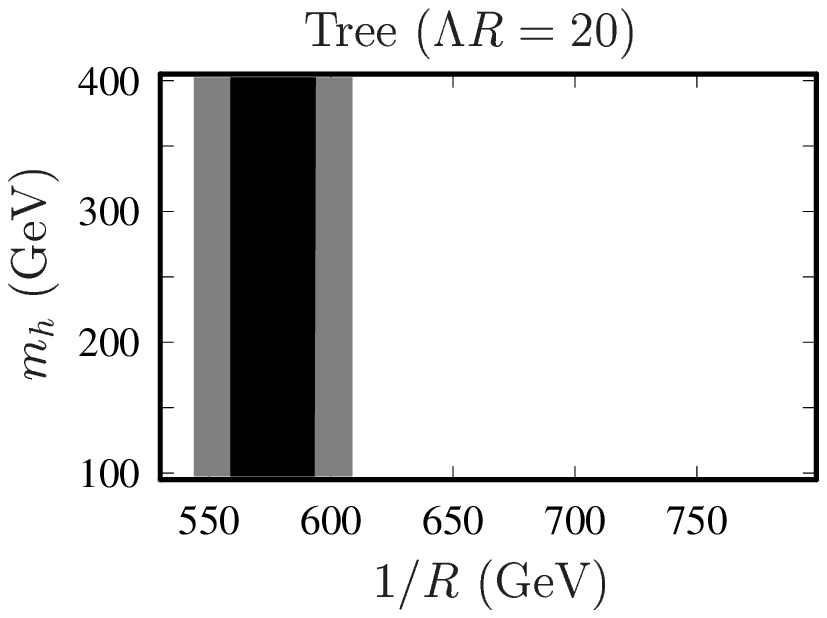}}
    \put(-100,-15){(c)}
    \hspace{0.5cm}
    \scalebox{0.80}{\includegraphics*{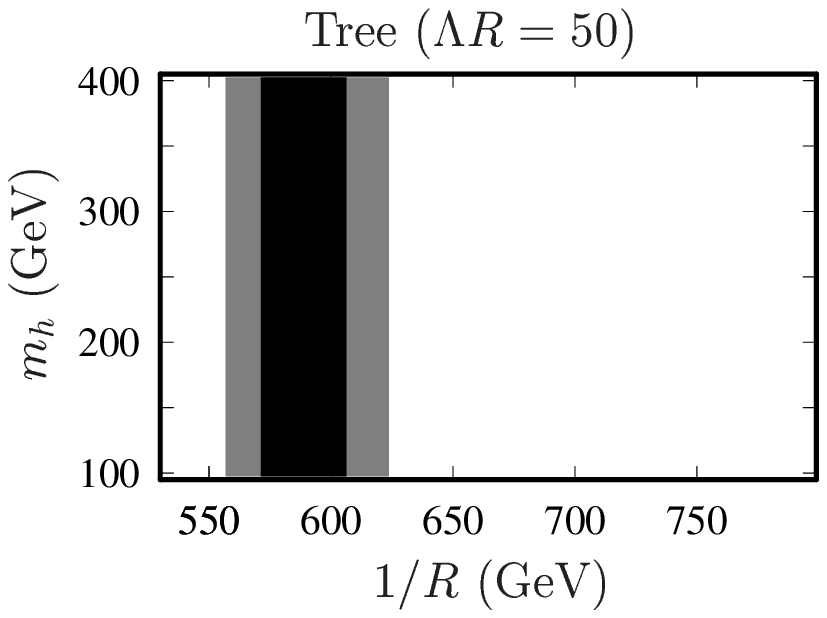}}
    \put(-100,-15){(d)}
    \caption{\footnotesize 
      Parameter region of the minimal UED model consistent
      with the WMAP observation in the ($1/R$, $m_h$) plane for
      $\Lambda R = 20$ (a) and for $\Lambda R = 50$ (b). 
      Black regions correspond to the 1$\sigma$ region of the 
      relic abundance measured by
      WMAP ($\Omega h^2 = 0.110 \pm 0.006$), and grey ones correspond
      to the 2$\sigma$ region.
      For comparison, tree-level results are also shown for
      $\Lambda R = 20$ (c) and for $\Lambda R = 50$ (d).}
    \label{fig:Region}
  \end{center}
\end{figure}

In Fig. \ref{fig:Region}, 
the parameter region of the minimal UED model consistent
with the WMAP observation is shown in the ($1/R, m_h$) plane. 
Black regions
correspond to the 1$\sigma$ region of the relic abundance measured 
by WMAP ($\Omega h^2 = 0.110 \pm 0.006$), and grey ones 
correspond to the 2$\sigma$ region. 
Here we take the cutoff scale to be $\Lambda R = 20$
(Fig. \ref{fig:Region}(a)) and $\Lambda R = 50$ 
(Fig. \ref{fig:Region}(b)).
For comparison, tree-level results are also shown
for $\Lambda R = 20$ (Fig. \ref{fig:Region}(c))
and for $\Lambda R = 50$ (Fig. \ref{fig:Region}(d)).

A comparison of Figs. \ref{fig:Region}(a) with (c)
shows that the compactification scale
$1/R$ consistent with the observation is increased by $5$ \% due to
the resonance.
In the allowed region, the mass difference between the second KK Higgs
and two LKPs is not so small, 
as seen in Fig. \ref{fig:degeneracy}(a). 
Therefore,
the resonance effect does not significantly alter the tree level result.

However, the resonance drastically changes the relic
abundance when $\Lambda R = 50$ as indicated
in Figs. \ref{fig:Region}(b) and (d).
The allowed region is strongly sensitive to the mass of the second KK Higgs,
thus sensitive to the SM Higgs mass.
For $250$ GeV $\lesssim m_h \lesssim$ $300$ GeV,
due to the ``late time decreasing'' 
illustrated in Fig. \ref{fig:T-evolve}(b),
the allowed range of the compactification scale is extended
to a higher mass scale,
in sharp contrast to the preceding work \cite{Servant:2002aq}.
On the contrary, for 350 GeV $\lesssim m_h \lesssim$ 500 GeV,
the compactification scale is tightly constrained.
Notice that the narrow region agrees with the degeneracy line 
($\delta =0$) in Fig. \ref{fig:degeneracy}(b).
For $m_h \gg 500$ GeV, the second KK Higgs mass is
smaller than twice the LKP mass in the allowed region so that the resonance
disappears: 
the region coincides with those obtained by the tree-level calculation.


\vspace{1.0cm}
\lromn 4 \hspace{0.2cm} {\bf Summary and discussion}
\vspace{0.5cm}

In this work, 
we have evaluated the thermal relic abundance of the LKP dark matter
including not only the coannihilation but also the resonance
in the minimal UED model.
We have systematically investigated the effects of resonances on
each annihilation process. We found that only the resonance with the second KK Higgs
contributes to the abundance in the self-annihilation of LKP dark matters.

Furthermore, we have pointed out that the LKP dark matter abundance
strongly depends on both the SM Higgs boson mass $m_h$ 
and the cutoff scale $\Lambda$ due to the resonance.
A wider range of the compactification scale turns out to be
consistent with cosmological observations.
For example, $580$ GeV $\lesssim 1/R \lesssim$ $770$ GeV 
for $m_h \simeq 300$ GeV and $\Lambda R = 50$.
There also exist parameter region 
where the LKP mass is tightly constrained.
The non-trivial dependence originates from the fact that 
the second KK Higgs mass plays a crucial role in
determining the relic dark matter abundance due to
the resonant annihilation.


\section*{Acknowledgments}
The work of M.K. is supported in part by
the Japan Society for the Promotion of Science.


\end{document}